# Identification of the superconductivity in bilayer nickelate La$_3$Ni$_2$O$_7$ upon 100 GPa

Jingyuan Li[1,*], Di Peng[2,*], Peiyue Ma[1], Hengyuan Zhang[1], Zhenfang Xing[3], Xing Huang[1], Chaoxin Huang[1], Mengwu Huo[1], Deyuan Hu[1], Zixian Dong[1], Xiang Chen[1], Tao Xie[1], Hongliang Dong[2,3], Hualei Sun[4, †], Qiaoshi Zeng[2,3, ‡], Ho-kwang Mao[2,3], Meng Wang[1,§]

[1]*Center for Neutron Science and Technology, Guangdong Provincial Key Laboratory of Magnetoelectric Physics and Devices, School of Physics, Sun Yat-Sen University, Guangzhou, Guangdong 510275, China*
[2]*Shanghai Key Laboratory of Material Frontiers Research in Extreme Environments (MFree), Institute for Shanghai Advanced Research in Physical Sciences (SHARPS), Shanghai 201203, China*
[3]*Center for High Pressure Science & Technology Advanced Research, Shanghai, 201203, China*
[4]*School of Science, Sun Yat-Sen University, Shenzhen, Guangdong 518107, China*
[*]These authors contributed equally to this work.
[†] sunhlei@mail.sysu.edu.cn;
[‡] zengqs@hpstar.ac.cn;
[§]wangmeng5@mail.sysu.edu.cn

**Abstract:** Identification of superconductivity in the Ruddlesden-Popper phases of nickelates under high pressure remains challenging. Here, we report a comprehensive study of the crystal structure, resistance, and Meissner effect in single crystals of La$_3$Ni$_2$O$_7$ with hydrostatic pressures up to 104 GPa. X-ray diffraction measurements reveal a structural transition from the orthorhombic to a tetragonal phase above 40 GPa. Zero resistance of the superconductivity was achieved with a maximum onset $T_c^{onset}$ of 83 K at 18.0 GPa. Superconductivity is gradually suppressed until it disappears above 80 GPa, resulting in a right-triangle-like superconducting region. The direct-current magnetic susceptibility technique successfully detected the Meissner effect in La$_3$Ni$_2$O$_7$ under pressure; the maximum superconducting volume fraction is estimated to be 62.7% at 22.0 GPa. Thus, we demonstrate the bulk nature of superconductivity in the bilayer nickelate La$_3$Ni$_2$O$_7$ single crystals under high pressure. The results reveal intimate connections among the superconductivity, oxygen content, and structure in La$_3$Ni$_2$O$_7$.

Seeking unconventional superconductivity in nickel-oxide materials has been a long-lasting topic due to the similar lattice and electronic structures with cuprates. Especially the Ni$^+$ in $Re$NiO$_2$ ($Re$ = La, Nd, Sm, etc.) has the same spin configuration as Cu$^{2+}$ in cuprates [1,2]. However, superconductivity wasn't realized in nickelates until the superconductivity with a transition temperature of $T_c$ = 15 K was reported in Nd$_{0.8}$Sr$_{0.2}$NiO$_2$ thin film samples in 2019 [3–5]. Ni ions with a valence state close to Ni$^+$ and a spin S=1/2 in planar coordination with oxygen ions were thought crucial for the emergence of superconductivity in nickelates [2,6]. The discovery of superconductivity in the Ruddlesden-Popper (RP) phase nickelate $La_3Ni_2O_7$ with $T_c \sim$ 80 K has ignited renewed interest in the field of condensed matter physics [7–12]. The superconductivity was originally reported in the bilayer RP phase of $La_3Ni_2O_7$ with a '2222' stacking sequence of the $NiO_6$ octahedra between 14.0-43.5 GPa. The bilayer phase was suggested to undergo a structural transition from a low-pressure orthorhombic *Amam* phase to a high-pressure orthorhombic *Fmmm* phase, where the bond angle of Ni-O-Ni along the *c* axis changes from 168° to 180° [7]. Scanning transmission electron microscopy (STEM) [13] and x-ray diffraction (XRD) measurements [14] confirmed the '2222' stacking sequence on single crystals of $La_3Ni_2O_7$ grown by the high-pressure floating zone furnace and polycrystalline samples of $La_3Ni_2O_7$ [15] and $La_2PrNi_2O_7$ [16] grown by the sol-gel method. Oxygen vacancies were visualized on the inner apical oxygen site shared by two $NiO_6$ octahedra [13]. Further structural analysis under high pressure and low temperature revealed a tetragonal *I*4/*mmm* phase corresponding to the compressed superconducting (SC) state [17]. Recently, a new structure of $La_3Ni_2O_7$ with an alternating monolayer-trilayer $NiO_6$ octahedra stacking sequence, denoted as the '1313' phase, was identified [18–21]. It is unclear how the distinct structures might promote the high-$T_c$ superconductivity in $La_3Ni_2O_7$.

The other enigmatic phenomena are the difficulty of realizing zero resistance and weakly suppressed $T_c$ in $La_3Ni_2O_7$ under pressures below 43.5 GPa [7,9–11]. Low SC volume fraction revealed by the alternating-current (*ac*) magnetic susceptibility techniques and poor repeatability question the reliability of bulk superconductivity [7,11,15,19]. Recently, both zero resistance and the Meissner effect from direct-current (*dc*) magnetic susceptibility in the trilayer nickelate $La_4Ni_3O_{10}$ under pressure were reported, yielding a bulk superconductivity with $T_c^{onset} \sim 30$ K [12,22–25]. Accordingly, it is important to examine the nature of the superconductivity in the bilayer nickelate $La_3Ni_2O_7$.

In this work, we reinvestigate the high-pressure structure of $La_3Ni_2O_7$, explore the electric properties, and clarify the controversy of bulk or filamentary superconductivity in compressed $La_3Ni_2O_7$ single crystals. By employing neon gas as the pressure transmitting medium, in addition to the previously reported *Amam* to *Fmmm* structural transition, we show an orthorhombic to tetragonal transition above 40 GPa at room temperature. The tetragonal phase persists to 100 GPa, which is the maximum pressure in our XRD measurements. Furthermore, the high-pressure transport measurements up to 104 GPa reveal a phase diagram of pressure-driven superconductivity with a maximum $T_c^{onset}$ of 83 K. The SC phase is virtually suppressed above 80 GPa. The maximum SC volume fraction is estimated up to ~62.7% by measuring the high-pressure $dc$ magnetic susceptibility, demonstrating bulk superconductivity in $La_3Ni_2O_7$.

The high-pressure synchrotron XRD measurements were conducted at the BL15U1 beamline at the Shanghai Synchrotron Radiation Facility (SSRF) with a wavelength of $\lambda = 0.6199$ Å. Polycrystal samples synthesized by the sol-gel method were used in the XRD measurements [15]. The diamond anvil cell (DAC) electric transport and $dc$ magnetic susceptibility measurements were performed on single crystals of $La_3Ni_2O_7$ from the same batch we investigated before [7,14,26]. Both the polycrystal and single-crystal samples were confirmed to be the bilayer structural RP phase (see supplementary Fig. S1). Details of the oxygen content analysis have been discussed elsewhere and are beyond the scope of this work [13]; we thus use $La_3Ni_2O_7$ to represent the composition of our samples for simplicity in this work. A custom-designed miniature DAC made of beryllium-copper alloy was employed to conduct ultrasensitive $dc$ magnetic susceptibility measurements. The $dc$ measurements were executed utilizing the magnetic property measurement system (MPMS, Quantum Design).

Our previous synchrotron XRD results indicate a structural transition from the *Amam* to the *Fmmm* space group at ~14 GPa [7]. Silicon oil was employed as the pressure-transmitting medium (PTM). The pressure inhomogeneity induced by the solidification at 3 GPa and the glass-to-glass transitions of silicon oil at 10 and 16 GPa may involve an artificial effect on the data analysis [27]. In this high-pressure XRD measurements on polycrystalline $La_3Ni_2O_7$ samples up to 100 GPa, neon gas was used as the PTM. An anomaly can be observed from the relative change of the diffraction peaks against pressure at 12.3 GPa (see supplementary Fig. S2), which is well above the crystallization pressure of 4.8 GPa and away from the non-hydrostatic pressure of 16.0 GPa of neon [28]. Figure 1b shows the peak widths of (115) and (020)/(200). The width of the diffraction peak (115) is broadened gradually with increasing pressure. Due to broadening by the instrumental resolution and small orthorhombicity, we fit (020) and (200) as one peak. The width of (020)/(200) is wider than that of (115) below 12.3 GPa. The differences extend up to 40 GPa, above where (020) and (200) could not be distinguished from each other, indicating it is the tetragonal structure (space group *I4/mmm*) within the instrumental resolution. The peak width of (020)/(200) also shows a kink at 12.3 GPa, which may correspond to the *Amam* to the *Fmmm* space group transition. Figure 1c and 1d plot the refined lattice parameters.

The superconductivity in La$_3$Ni$_2$O$_7$ emerged from either a weakly insulating state or a metallic state at ambient pressure [7–11]. The $T_c$ is robust below 43.5 GPa. It is important to elucidate the effect of pressure on superconductivity from non-SC phase to ultrahigh pressure. To justify the reliability of experimental results, we conducted high-pressure electric measurements on five samples with a similar size of 30×30×10 μm$^3$ from the same batch, as shown in Fig. 2. KBr was adopted as the PTM in our transport measurements. Figure 2a shows the resistance at pressures from 0 to 104 GPa. At ambient pressure, the resistance shows a metallic behavior and an anomaly at $T^*$~140 K, which may be related to a density-wave transition [14,26,29–32]. By applying 0.9 GPa pressure in the DAC in run 1, the anomaly in resistance cannot be observed. A drop in resistance occurs at 8 K and 10.6 GPa, indicating the emergence of superconductivity. The maximum $T_c^{onset}$ is 73 K at 21.4 GPa in run1, while a residual resistance of 1 mΩ remains at 2 K. Superconductivity is suppressed gradually until the onset transition cannot be identified at 80.2 GPa and higher pressures. Figures 2b-2e and S2 show the high-pressure transport measurements on the other samples. The maximum $T_c^{onset}$ reaches 83 K at 18.2 GPa in run 2. Zero resistance is observed with the solid KBr as the PTM in run 2, run 3, and run 4, as shown in Fig. 2d, indicating a good sample quality. Figures 2f and 2g show suppression of superconductivity by a magnetic field. A Ginzberg-Landau fitting of the upper critical field is 126 T for run 4.

To further investigate the nature of the superconductivity of La$_3$Ni$_2$O$_7$ under pressure, a high-pressure *dc* magnetic susceptibility measurement was employed to measure the Meissner effect. The DAC utilized in this experiment featured a pair of diamond anvils, each with a diameter of 400 μm. Non-magnetic rhenium gaskets were adopted to minimize the interference of the magnetic measurements. The single crystal sample of La$_3$Ni$_2$O$_7$ was carefully positioned within the sample chamber and had dimensions of approximately 180 μm in diameter and 20 μm in thickness. Helium gas was selected as the PTM to achieve optimal hydrostatic pressure conditions. Helium is well-suited for this purpose due to its low viscosity and high compressibility, facilitating uniform pressure transmission to the sample.

Distinct from the nearly constant background signals in Fig. 3a, significant diamagnetic behaviors around $T_c^{onset}$ ~ 76 K emerged under 22.0 GPa in both zero field cooling (ZFC) and field cooling (FC) measurements in run1 (Fig. 3b). The SC volume fraction estimated at 67 K is around 16% and 13.6% for ZFC and FC, respectively. A larger volume fraction, around 62.7%, is yielded down to 20 K. Details of the method for estimating the SC volume fraction could be found elsewhere [22]. Such diamagnetic behavior is reproducible in *dc* magnetic measurements in run 2 and run 3 under 20.0 and 21.0 GPa (Figs. 4c-4d), yielding SC volume fractions of 42.1% and 46.3%, respectively. The notable *dc* diamagnetic signals detected in pressurized $La_3Ni_2O_7$ for the first time elucidate the Meissner effect and demonstrate the bulk nature of the superconductivity of $La_3Ni_2O_7$.

The results of the identification of superconductivity measured under high pressure are summarized in Fig. 4. The $T_c^{onset}$ is defined as the onset SC transition temperature, while the $T_c^{mid}$ is defined as the temperature where the resistance is in the middle of that at the $T_c^{onset}$ and 2 K. The background color scale is the relative resistance change to that at 150 K in run 1. Superconductivity emerges roughly coincidently with the structural transition from *Amam* to *Fmmm*. The reasons may be due to a Fermi surface reconstruction or the interlayer magnetic exchange coupling enhancement as the structural transition [33–44]. The $T_c^{onset}$ abruptly reaches a maximum of 83 K at 18.0 GPa and then decreases with increasing pressure, outlining a right-triangle-like feature of the superconductivity in $La_3Ni_2O_7$ under pressure. The $T_c^{onset}$ drops to 38 K at 80.2 GPa. These features indicate a robust bulk SC phase in $La_3Ni_2O_7$.

The emergence of superconductivity in $La_3Ni_2O_7$ under high pressure exhibits sample and pressure homogeneity dependence [7–9,11]. Both zero resistance and *dc* diamagnetic susceptibility have been achieved in single crystals of $La_3Ni_2O_7$. However, the zero resistance can only be observed for small samples with typical sizes of $30\times30\times10$ μm$^3$ in our measurements. The SC volume fraction estimated from the *dc* magnetic susceptibility measurements varies from the previously obtained several percent to the maximum of 62.7% in this work, likely due to the oxygen vacancy and pressure inhomogeneity. This conjecture is supported by the STEM measurements, which directly reveal various distributions of the oxygen vacancies, especially at the inner apical oxygen site [13]. The inner apical oxygen is directly involved in the superexchange magnetic interactions of the two nickel ions along the *c*-axis; it affects the splitting of the bonding and antibonding states of the $3d_{z^2}$ orbitals [7,39,45–48]. According to extensive theoretical analysis, interlayer coupling plays an important role in the superconductivity of pressurized $La_3Ni_2O_7$ [33,34,36,42–44,49–52]. Inelastic neutron scattering [53] and resonance inelastic x-ray scattering measurements [31] indeed reveal a strong interlayer coupling in $La_3Ni_2O_7$ compared to the dominant intralayer couplings in copper-based [54] and iron-based superconductors [55]. It is reasonable to assume that the inner apical oxygen vacancies will suppress the superconductivity in the bilayer nickelate under pressure.

Compared to $La_2NiO_4$ and $La_4Ni_3O_{10}$, the bilayer $La_3Ni_2O_7$ is a metastable phase with a narrower oxygen pressure window of 10-18 bar during the single crystal growth [56]. This complicates the sample synthesis, and sample inhomogeneity is hard to avoid. The other intergrowth phases, such as $La_2NiO_4$, $La_4Ni_3O_{10}$, and some other stacking sequences, are also possible [18,21,57]. The SC volume fraction can be low for samples with more oxygen vacancies, and the superconductivity behaves like filamentary. However, zero resistance and Meissner effect measured in this work demonstrate that samples with a large SC volume fraction and zero resistance are achievable. The intimate relationship between pressure homogeneity and superconductivity in $La_3Ni_2O_7$ may be ascribed to the crucial role of the Ni-O-Ni bond angle in the SC state and the bilayer structure. The SC phase diagram suggests the optimal pressure, where $T_c^{onset}$ is the highest, is close to the pressure of the structural transition from *Amam* to *Fmmm*.

In conclusion, we have measured the structures and superconductivity of $La_3Ni_2O_7$ from ambient pressure to 104 GPa. The phase diagram displays a SC region that emerges with a structural transition and ends up around 80 GPa, different from that of carrier-doped or pressure-induced dome-shaped SC phase diagram in both cuprates and iron pnictides. Moreover, the Meissner effect of pressurized $La_3Ni_2O_7$ is detected. The calculated maximum SC volume fraction is 62.7% at 22.0 GPa, suggesting the bulk nature of superconductivity. These findings provide important insights into the current confusion about the superconductivity in compressed $La_3Ni_2O_7$.

Work at Sun Yat-Sen University was supported by the National Natural Science Foundation of China (Grants No. 12425404, No. 12494591, No. 12474137, No. 12174454, and No. 12304187), the National Key Research and Development Program of China (Grants No. 2023YFA1406000 and No. 2023YFA1406500 ), the Guangdong Basic and Applied Basic Research Foundation (Grants No. 2024B1515020040 and No. 2024A1515030030), the Guangzhou Basic and Applied Basic Research Funds (Grants No. 2024A04J6417 and No. 2024A04J4024), the Shenzhen Science and Technology Program (Grant No. RCYX20231211090245050), the Guangdong Provincial Key Laboratory of Magnetoelectric Physics and Devices (No. 2022B1212010008), and the Research Center for Magnetoelectric Physics of Guangdong Province (No. 2024B0303390001). D.P., Q.Z., and H.M. acknowledge the support from Shanghai Key Laboratory of Material Frontiers Research in Extreme Environments, China (No. 22dz2260800), the Shanghai Science and Technology Committee, China (No. 22JC1410300). We also thank the BL15U1 station and User Experiment Assist System in Shanghai Synchrotron Radiation Facility (SSRF) for the help in high-pressure structural characterizations.

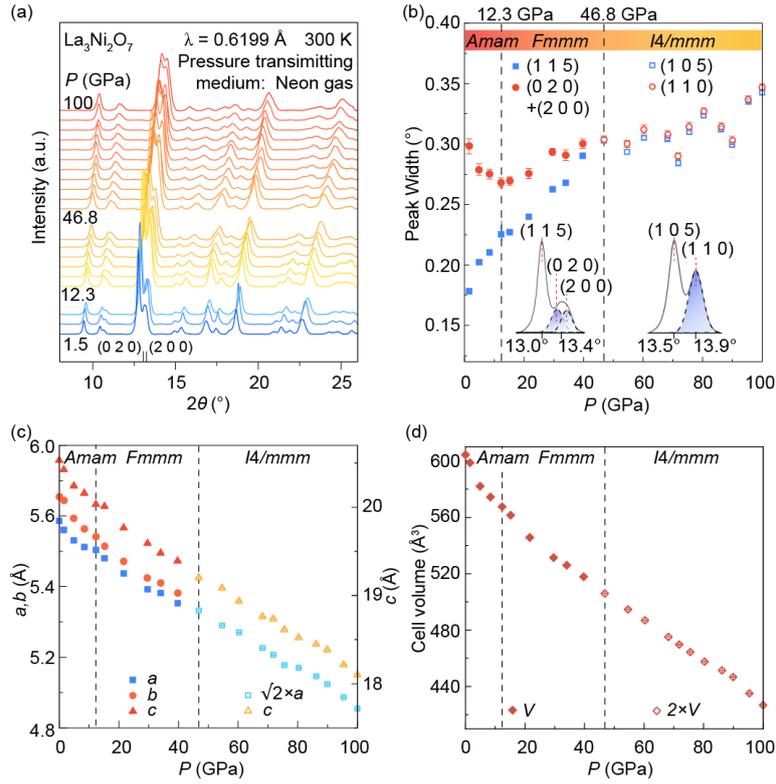

FIG. 1. High-pressure structural characterizations of $La_3Ni_2O_7$ up to 100 GPa. (a) Synchrotron XRD patterns of $La_3Ni_2O_7$ under pressures from 1.5 to 100 GPa. The pressure transmitting medium (PTM) was neon gas. (b) Pressure dependence of the peak widths of (0 2 0)/(2 0 0) as a whole and (1 1 5) in the *Amam* space group. A merging of the peak widths at 12.3 GPa signals a structural transition from the *Amam* phase to the *Fmmm* phase. The peak widths of (0 2 0)/(2 0 0) become indistinguishable from those of (1 1 5) above 46.8 GPa, indicating a structural transition from *Fmmm* to tetragonal *I4/mmm* space group. The reflection indexes of (1 1 5) and (0 2 0)/(2 0 0) in the *Amam* space group change to (1 0 5) and (1 1 0) in the *I4/mmm* space group. The insets are the zoom-in experimental data. (c) Lattice parameters of $La_3Ni_2O_7$ obtained from the synchrotron XRD data. (d) Cell volumes of $La_3Ni_2O_7$.

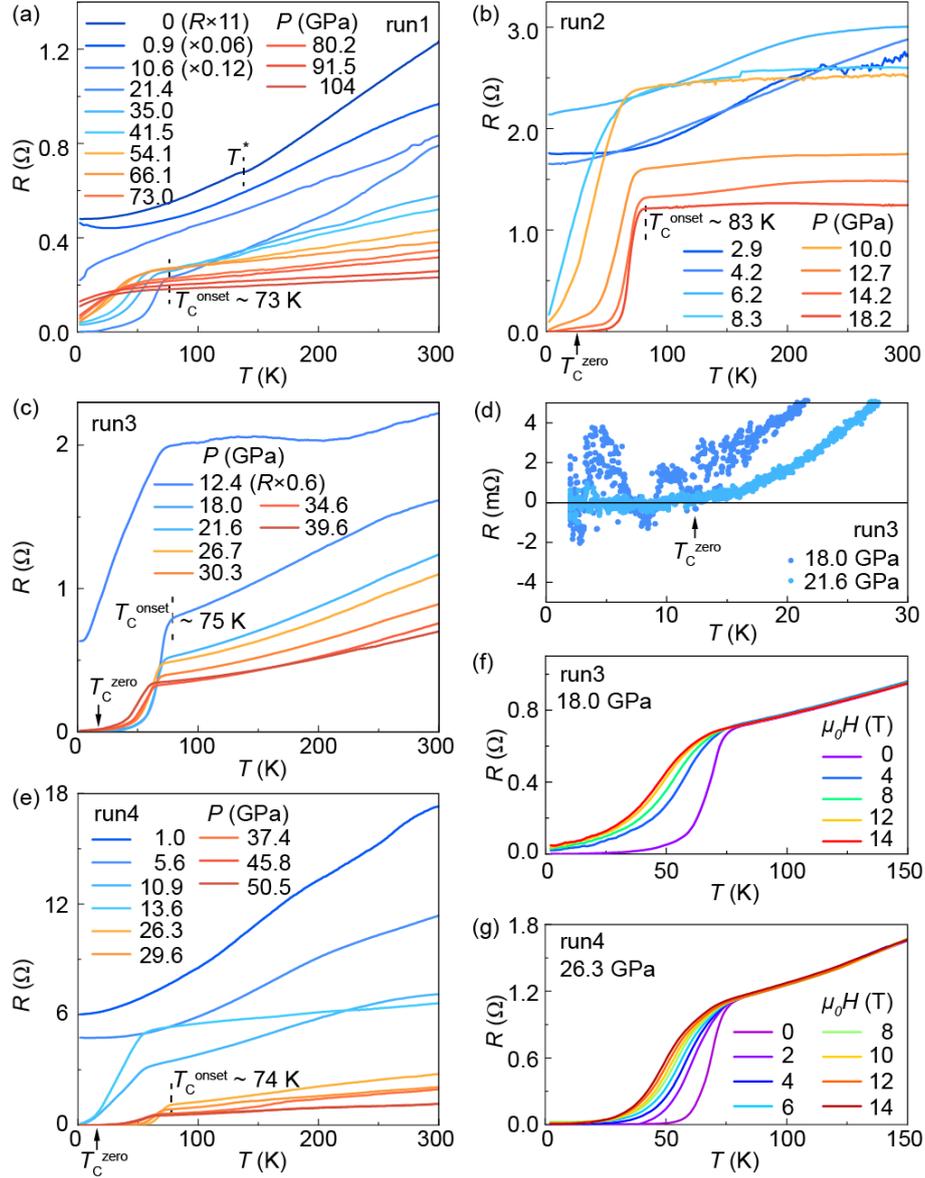

FIG. 2. Temperature dependence of the in-plane resistance of La$_3$Ni$_2$O$_7$ under various pressures. (a)-(c), (e) High-pressure resistance curves from run 1 to run 4. The resistance of La$_3$Ni$_2$O$_7$ from ambient pressure to 104 GPa is measured in run 1. A $T_c^{onset}$ ~ 83 K is observed in run 2. (d) A zoom-in view of the resistance curves of run 3 under 18.0 and 21.6 GPa below 30 K. Zero resistance is achieved. (f)-(g) Field-dependent resistance curves at 18.0 GPa of run 3 and 26.3 GPa of run 4.

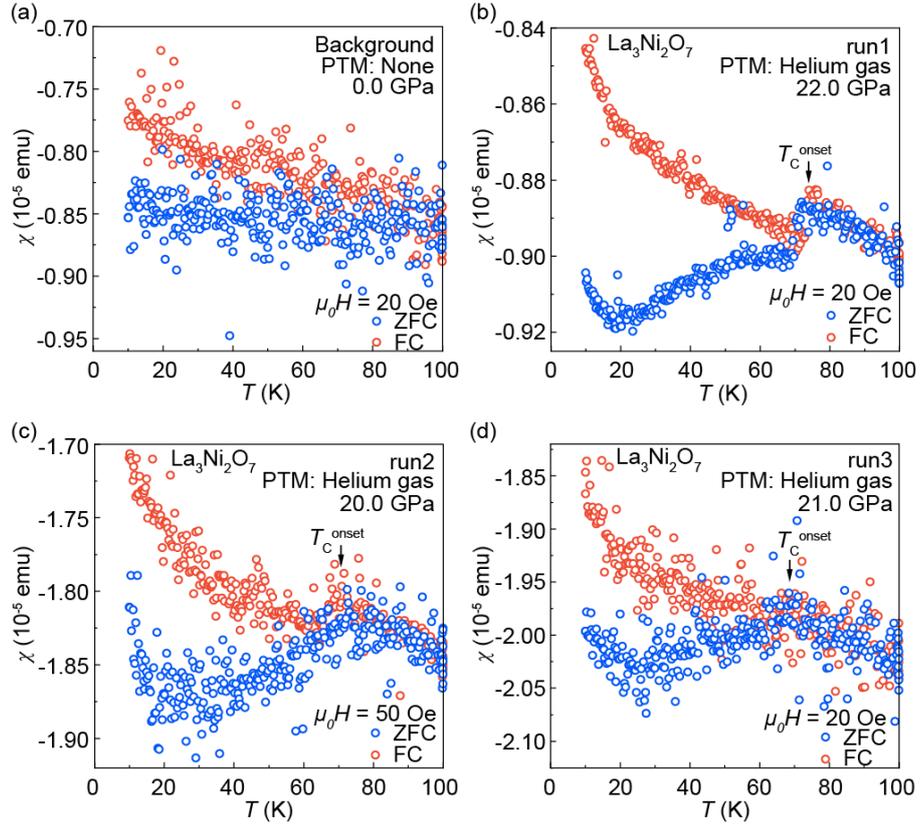

FIG. 3. Direct current magnetic susceptibility measurements of $La_3Ni_2O_7$ under pressure. Zero-field cooling (ZFC) and field cooling (FC) curves are measured with a field perpendicular to the *ab* plane. (a) Background signals of the high-pressure cell without a sample at ambient pressure. (b)-(d) Evident Meissner effect induced diamagnetic signals of pressurized $La_3Ni_2O_7$ single crystal from run 1 to run 3. Both ZFC and FC curves show prominent diamagnetic responses. The black arrows indicate the onset temperatures of superconductivity.

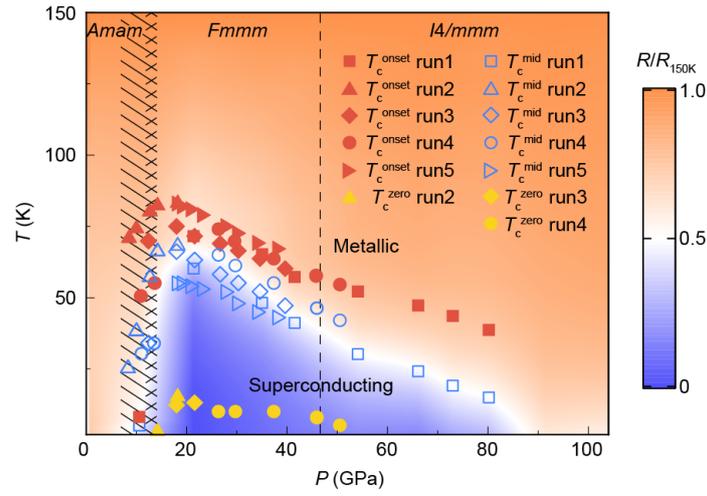

FIG. 4. The superconducting phase diagram of $La_3Ni_2O_7$ single crystals under ambient pressure to 104 GPa. The red solid symbols represent the onset temperatures of superconductivity $T_c^{onset}$ obtained from five runs. The blue hollow symbols represent the middle temperatures $T_c^{mid}$ of the SC transition defined by the temperature corresponding to the resistance of $R_{mid}=(R_{onset}+R_{2K})/2$. The yellow dots mark the zero resistance temperatures $T_c^{zero}$ of run 2 to run 4. The color of the ground shows the data of run 1. The structural transition pressure is indicated by the black stripe lines and the dashed line.

# Supplementary Information : Identification of the superconductivity in bilayer nickelate La$_3$Ni$_2$O$_7$ upon 100 GPa

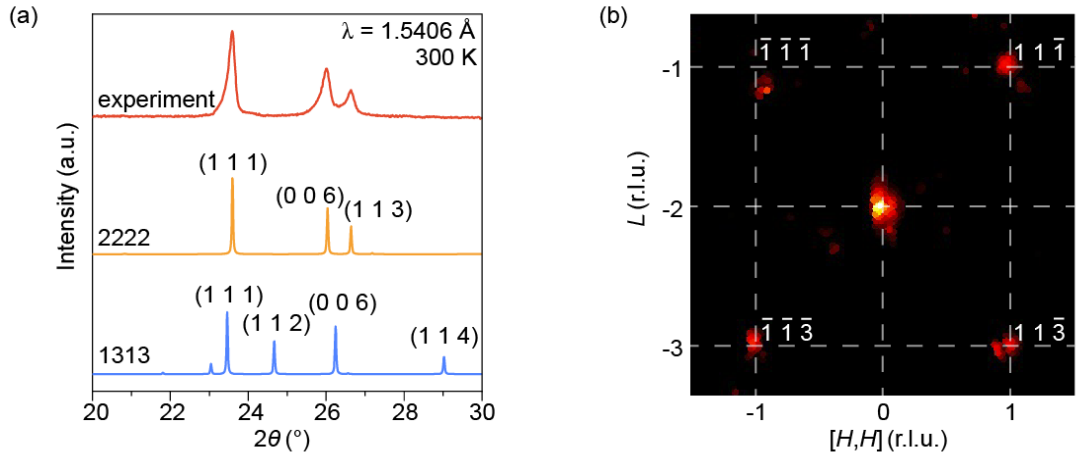

FIG. S1. Structural characterizations of La$_3$Ni$_2$O$_7$ at ambient pressure. (a) Powder X-ray diffraction (XRD) pattern at ambient pressure (red line). The simulated XRD patterns of the 2222-phase (orange line) and 1313-phase (blue line) are shown below for comparison. (b) A slice of single crystal XRD pattern in the (*H H L*) plane. The (1 1 3) peak presents while the (1 1 2) peak is absent, consistent with the '2222' bilayer structure.

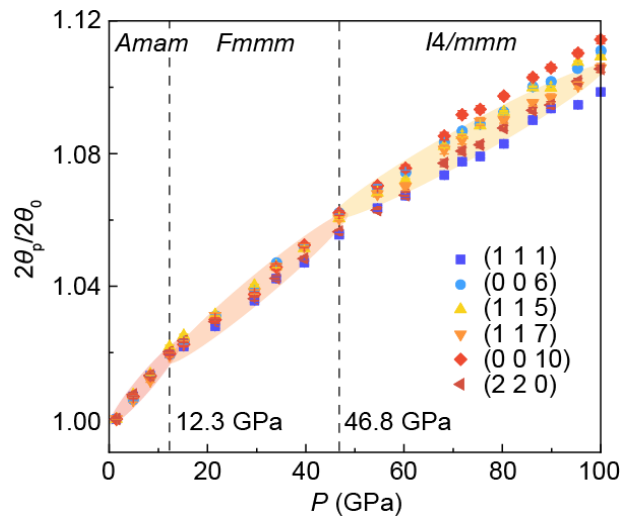

FIG. S2. Pressure dependence of the peak positions of six isolated Bragg peaks. The peaks are labeled by the Miller indices in the *Amam* space group.

**TABLE S1** Crystal parameters of pressurized La$_3$Ni$_2$O$_7$ obtained from the Rietveld refinements.

| Pressure (GPa) | $a$ (Å) | $b$ (Å) | $c$ (Å) | Space group | $R_{wp}$ (%) | $R_p$ (%) | GOF |
|---|---|---|---|---|---|---|---|
| 0 | 5.3964(2) | 5.4549(1) | 20.5293(8) | *Amam* | 4.30 | 6.50 | 2.51 |
| 1.5 | 5.3605(8) | 5.4448(8) | 20.424(4) | *Amam* | 6.70 | 9.40 | 0.86 |
| 4.9 | 5.3309(7) | 5.3939(7) | 20.242(3) | *Amam* | 9.31 | 10.52 | 0.90 |
| 8.4 | 5.3123(6) | 5.3631(7) | 20.159(3) | *Amam* | 8.63 | 10.10 | 0.89 |
| 12.3 | 5.3041(9) | 5.3417(8) | 20.031(4) | *Fmmm* | 10.47 | 13.80 | 1.20 |
| 15.2 | 5.2806(8) | 5.3144(9) | 20.010(4) | *Fmmm* | 10.02 | 12.84 | 1.14 |
| 21.6 | 5.2376(7) | 5.2714(8) | 19.766(2) | *Fmmm* | 9.60 | 12.85 | 1.18 |
| 29.6 | 5.1925(9) | 5.2248(8) | 19.590(3) | *Fmmm* | 10.32 | 11.57 | 1.15 |
| 34.0 | 5.1820(7) | 5.2108(5) | 19.480(2) | *Fmmm* | 7.73 | 9.19 | 0.99 |
| 39.7 | 5.1531(8) | 5.1815(6) | 19.390(3) | *Fmmm* | 7.45 | 8.06 | 0.77 |
| 46.8 | 3.6296(5) | 3.6296(5) | 19.201(5) | *I4/mmm* | 8.64 | 9.46 | 0.84 |
| 54.6 | 3.6001(6) | 3.6001(6) | 19.083(3) | *I4/mmm* | 9.21 | 10.50 | 0.97 |
| 60.3 | 3.5858(6) | 3.5858(6) | 18.935(4) | *I4/mmm* | 9.10 | 11.60 | 0.91 |
| 68.2 | 3.5547(6) | 3.5547(6) | 18.761(4) | *I4/mmm* | 8.76 | 10.80 | 0.71 |
| 71.8 | 3.5408(8) | 3.5408(8) | 18.734(5) | *I4/mmm* | 10.90 | 13.20 | 1.04 |
| 75.5 | 3.5206(7) | 3.5206(7) | 18.613(4) | *I4/mmm* | 11.60 | 14.60 | 1.22 |
| 80.3 | 3.5148(6) | 3.5148(6) | 18.522(4) | *I4/mmm* | 10.00 | 12.00 | 0.96 |
| 86.2 | 3.4980(7) | 3.4980(7) | 18.447(4) | *I4/mmm* | 11.20 | 15.90 | 1.65 |
| 89.9 | 3.4822(9) | 3.4822(9) | 18.384(5) | *I4/mmm* | 12.20 | 14.90 | 1.45 |
| 95.4 | 3.4560(8) | 3.4560(8) | 18.216(5) | *I4/mmm* | 11.90 | 15.20 | 0.98 |
| 100.0 | 3.433(1) | 3.433(1) | 18.101(6) | *I4/mmm* | 12.60 | 14.50 | 1.56 |

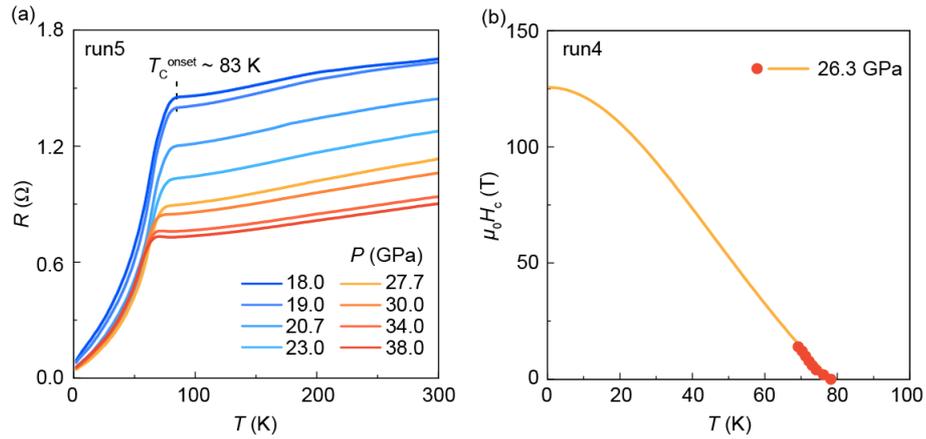

FIG. S3. (a) High-pressure resistance curves of run5. (b) The Ginzberg-Landau fitting of the upper critical field $\mu_0 H_{c2}$ of run4 under 26.3 GPa.